\documentclass[final,3p,twocolumn]{elsarticle}

\usepackage{amsmath}

\usepackage{graphicx}
\usepackage[caption=false]{subfig}

\usepackage{xcolor}
\usepackage{threeparttable}
\usepackage[hidelinks]{hyperref}

\usepackage{setspace}
\usepackage{lineno}

\usepackage{fontawesome5}

\journal{Nuclear Instruments and Methods A}

\begin{document}

\begin{frontmatter}

\title{Absolute light yield of the EJ-204 plastic scintillator}

\author[1]{J.~A.~Brown\corref{cor2}}
\cortext[cor2]{\copyright 2023. This manuscript version is made available under the CC BY-NC-ND 4.0 license \faCreativeCommons\ \faCreativeCommonsBy\ \faCreativeCommonsNc\ \faCreativeCommonsNd}
\author[1]{T.~A.~Laplace}
\author[1,2]{B.~L.~Goldblum\corref{cor1}}
\ead{bethany@nuc.berkeley.edu}
\cortext[cor1]{Corresponding author}
\author[1,3]{J.~J.~Manfredi}
\author[1]{T.~S.~Johnson} 
\author[2]{F.~Moretti} 
\author[1]{A.~Venkatraman}

\address[1]{Department of Nuclear Engineering, University of California, Berkeley, California 94720 USA}
\address[2]{Nuclear Science Division, Lawrence Berkeley National Laboratory, Berkeley, CA, 94720 USA}
\address[3]{Air Force Institute of Technology, Department of Engineering Physics, Wright-Patterson Air Force Base, OH 45433 USA}

\begin{abstract}
\begin{linenumbers}
The absolute light yield of a scintillator, defined as the number of scintillation photons produced per unit energy deposited, is a useful quantity for scintillator development, research, and applications. Yet, literature data on the absolute light yield of organic scintillators are limited. The goal of this work is to assess the suitability of the EJ-204 plastic scintillator from Eljen Technology to serve as a reference standard for measurements of the absolute light yield of organic scintillators. Four EJ-204 samples were examined: two manufactured approximately four months prior and stored in high-purity nitrogen, and two aged approximately eleven years and stored in ambient air. The scintillator response was measured using a large-area avalanche photodiode calibrated using low energy $\gamma$-ray and X-ray sources. The product of the quantum efficiency of the photodetector and light collection efficiency of the housing was characterized using an experimentally-benchmarked optical photon simulation. The average absolute light yield of the fresh samples, 9100 $\pm$ 400 photons per MeV, is lower than the manufacturer-reported value of 10400 photons per MeV. Moreover, the aged samples demonstrated significantly lower light yields, deviating from the manufacturer specification by as much as 26\%. These results are consistent with recent work showcasing environmental aging in plastic scintillators and suggest that experimenters should use caution when deploying plastic scintillators in photon counting applications.
\end{linenumbers}
\end{abstract}

                                             
\end{frontmatter}

\section{Introduction}
\label{sec:intro}

An understanding of the fundamental properties of organic scintillators is critical for advancing scintillation physics and enabling the use of these materials in basic science and applications. The absolute light yield, or number of scintillation photons produced per unit energy deposited, is an important property, as it impacts the temporal and energy resolution of scintillation detectors and is a critical input for optical photon transport models~\cite{Roncali2017, Kandemir2018, Moustafa2022} and maximum likelihood approaches to neutron image reconstruction~\cite{Weinfurther2018,Manfredi2020}. Given the experimental challenges associated with absolute light yield measurements, including requirements to quantify the emission spectrum of the scintillator, light collection efficiency of the housing, and quantum efficiency of the photodetector, the absolute light yield is often determined in ratio to a reference material. However, measurements of the absolute light yield of organic scintillators are scarce~\cite{Holl1988,Sysoeva2007}, which limits options for a reference standard and introduces challenges in metrological traceability.

Pioneering work by Holl et al.~\cite{Holl1988} provided measurements of the absolute light yield of two commercially available plastic scintillators, NE-110 (equivalent to EJ-208) and NE-102A (equivalent to EJ-212), obtaining 10,400 and 10,050 photons per MeV, respectively. However, Holl et al.\ characterized the light output of the scintillators using the half height of the Compton edge of a mono-energetic $\gamma$-ray spectrum (as opposed to fitting the Compton edge with a Monte Carlo simulation of the electron energy distribution convolved with the detection resolution), a known bias in organic scintillator light output calibration~\cite{Dietze1982}. In addition, the NE-110 absolute light yield measured by Holl et al.\ is inconsistent with the absolute light yield reported for the EJ-208 commercial equivalent (9,200 photons per MeV~\cite{Eljen-EJ-20x-specs}). Given these discrepancies and the dearth of data in the literature, additional measurements are required. 

The goal of this work is to assess the viability of using the EJ-204 plastic scintillator from Eljen Technology~\cite{Eljen-EJ-20x-specs} as a reference standard for measurements of the absolute light yield of organic scintillators. EJ-204 is suited for a broad range of radiation detection applications~\cite{Blain2017,AlHamrasdi2017,Manna2020,Erhart2022} due to its fast decay time, relatively high scintillation efficiency, and long optical attenuation length. It is particularly attractive for the detection of fast neutrons via collisions with protons within the hydrogenous medium. For example, this material was deployed as part of a neutron diagnostic system for fusion devices~\cite{Mitrani2021} and identified as a top candidate for compact kinematic neutron imagers currently under development~\cite{Sweany2019,Keefe2022,Cates2022}. The proton light yield of EJ-204 was previously measured over a broad energy range relative to the electron light output~\cite{Laplace2020}, and knowledge of the absolute light yield of the material would allow for determination of the number of optical photons produced by the recoil proton. However, the EJ-204 scintillator is composed of a polyvinyltoluene (PVT) base, which is known to exhibit yellowing~\cite{Barnaby1962} or become cloudy and ``fog''~\cite{Cameron2015} when subjected to environmental conditions, resulting in loss of light. While studies have been performed to examine the temperature dependence~\cite{Peralta2018} and radiation sensitivity~\cite{Li2005} of the optical properties of its commercial equivalent, BC-404, there is no published information on environmental aging of the EJ-204 scintillator. 

In this work, the absolute light yield of four EJ-204 samples was measured: two fresh samples manufactured approximately four months prior and stored in high-purity nitrogen, and two aged samples manufactured approximately eleven years prior and stored in ambient air. The experimental methods, including materials, instrumentation, and analysis procedures, are detailed in Section~\ref{sec:methods}. Section~\ref{sec:Results} provides the measured EJ-204 absolute light yield and a discussion of the results. A summary is provided in Section~\ref{sec:Summary}. 

\section{Experimental Methods}
\label{sec:methods}

\begin{figure}
	\center
	\includegraphics[width=0.4\textwidth]{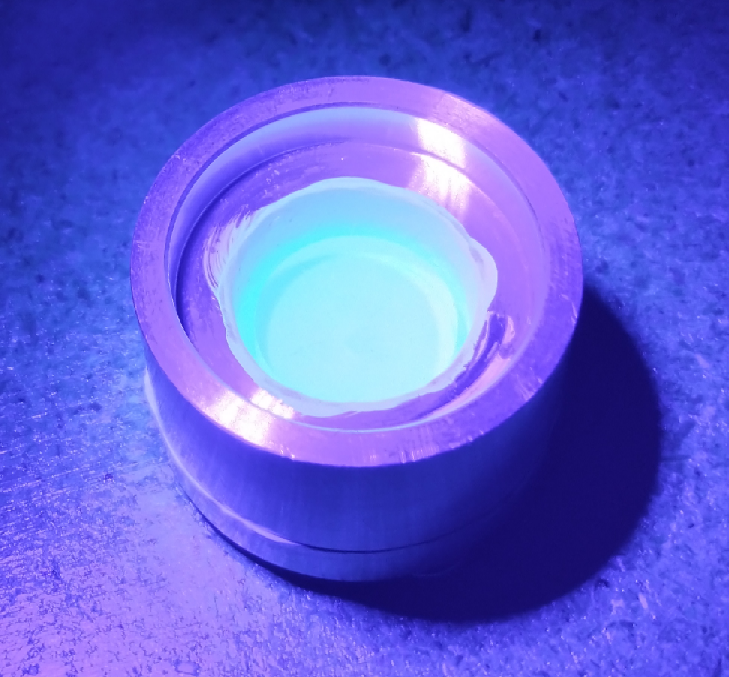}
	\caption{(Color online) EJ-204 scintillator in aluminum housing coated with Labsphere 6080 reflective paint. \label{fig:scintInCap}}
\end{figure}

Building upon prior work~\cite{Holl1988,deHaas2005,deHaas2008,Moszynski1997,Moszynski2000,Moszynski2002}, the absolute light yield of the EJ-204 scintillator was measured using the Compton edge of a 1274.5~keV $\gamma$ ray from a $^{22}$Na source and a large-area avalanche photodiode (APD) of Type SD 630-70-73-500 from Advanced Photonix, Inc.\, where the electron-hole (e-h) pair yield of the scintillation light was corrected for the spectrum-averaged quantum efficiency of the photodetector and light collection efficiency of the scintillator housing. That is, the absolute light yield, $L(E)$, produced by an electron of energy, $E$, is given by: 
\begin{equation}
\label{ALYgovern}
L(E) = \frac{N_{e{\text -}h}(E)}{ \int{P_s(\lambda) \eta(\lambda) \epsilon_{LC}(\lambda)} d\lambda} 
\end{equation}
where $N_{e{\text -}h}(E)$ is the e-h pair yield of the scintillation response, $P_s(\lambda)$ is the probability of scintillation emission with wavelength $\lambda$, $\eta(\lambda)$ is the quantum efficiency of the photodetector, and $\epsilon_{LC}(\lambda)$ is the light collection efficiency of the scintillator housing. 

The EJ-204 scintillator samples used in this study were 15-mm-dia.\ x 5-mm-h.\ right circular cylinders with diamond milled surface finish. To maximize light collection efficiency and maintain a reproducible geometry, the scintillator was placed in an aluminum housing internally coated with Labsphere~6080 reflective paint~\cite{labsphere}, shown in Fig.~\ref{fig:scintInCap}. The e-h pair yield of the scintillation response, $N_{e{\text -}h}(E)$, was determined relative to the e-h pair yield of low energy $\gamma$~rays and X~rays from $^{57}$Co and $^{109}$Cd sources detected directly in the photodiode. Measurements were performed to quantify each of the parameters in~\eqref{ALYgovern} and are discussed in further detail below.

\subsection{Emission Spectrum:  $P_s(\lambda)$}
\label{sec:emission}

The photoluminescence emission spectra of the EJ-204 samples were obtained using an Acton Research Corporation broad-spectrum XS-433 Xe lamp coupled to a SP-2150i monochromator. Additionally, X-ray luminescence spectra were obtained using a Nonius FR591 water-cooled rotating copper-anode X-ray generator (50~kV, 100~mA) from Bruker AXS Inc.\ as an excitation source. The emitted light was collected by a SpectraPro-2150i spectrometer with 50~$\mu$m exit slit and a PIXIS: 100B thermoelectrically-cooled charge-coupled device (CCD) (both from Princeton Instruments) with response in the $220-900$~nm range. The photoluminescence measurements were performed with a single grating (300 lines/mm, 300~nm blaze, centering at 350~nm), whereas the X-ray luminescence measurements employed dual gratings and order sorting filters~\cite{Derenzo2008}. In both cases, measurements were performed in a front face configuration (i.e., with the excitation source and detector facing the same side of the sample). Emission spectra were corrected to take into account the system response using the method of Derenzo et al.~\cite{Derenzo2008, Derenzo2018}. 

\begin{figure}
	\center
	\subfloat[Fresh sample \label{new-em}]{
	\includegraphics[width=0.47\textwidth]{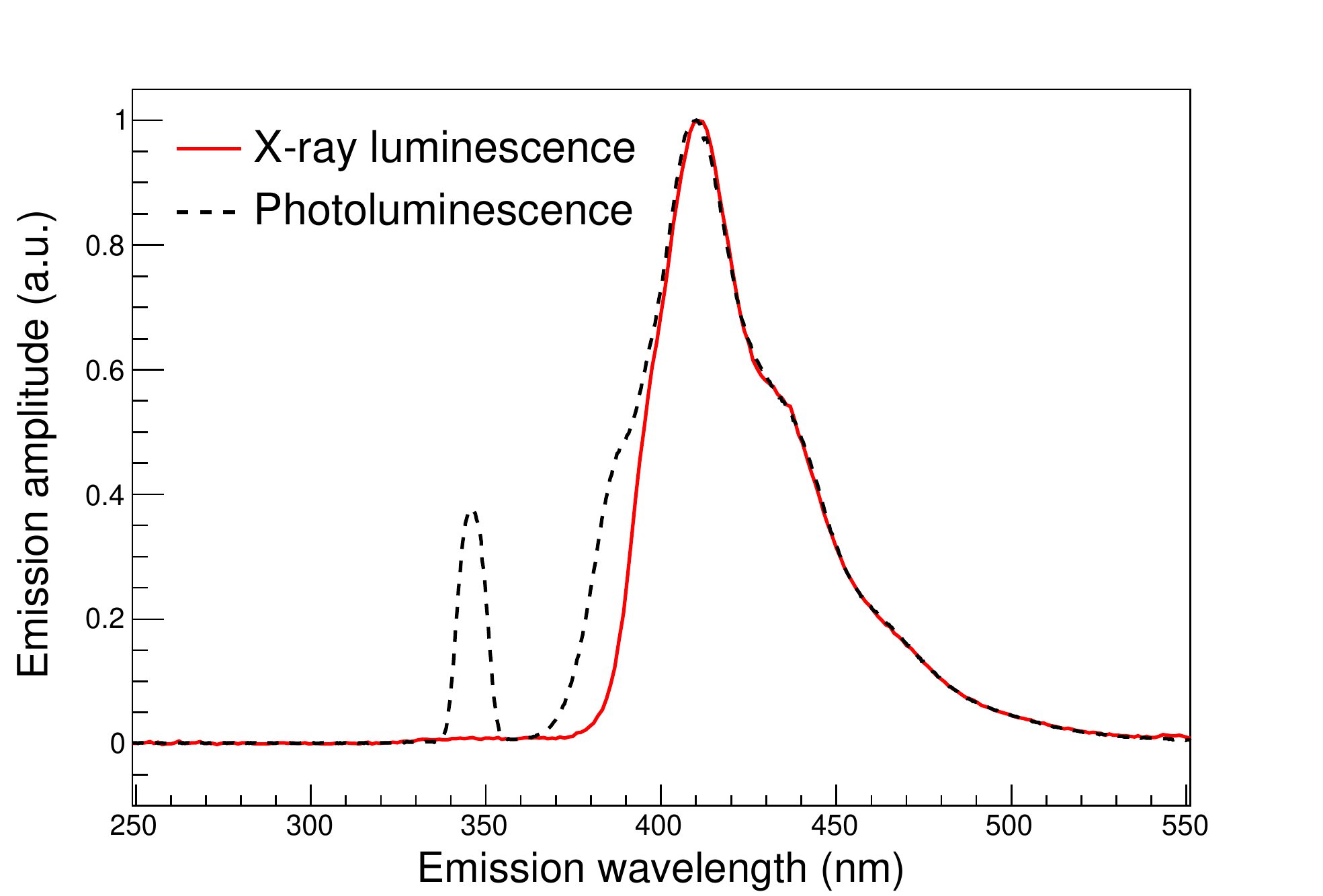}
	}\hfill
	\subfloat[Aged sample \label{old-em}]{
	\includegraphics[width=0.47\textwidth]{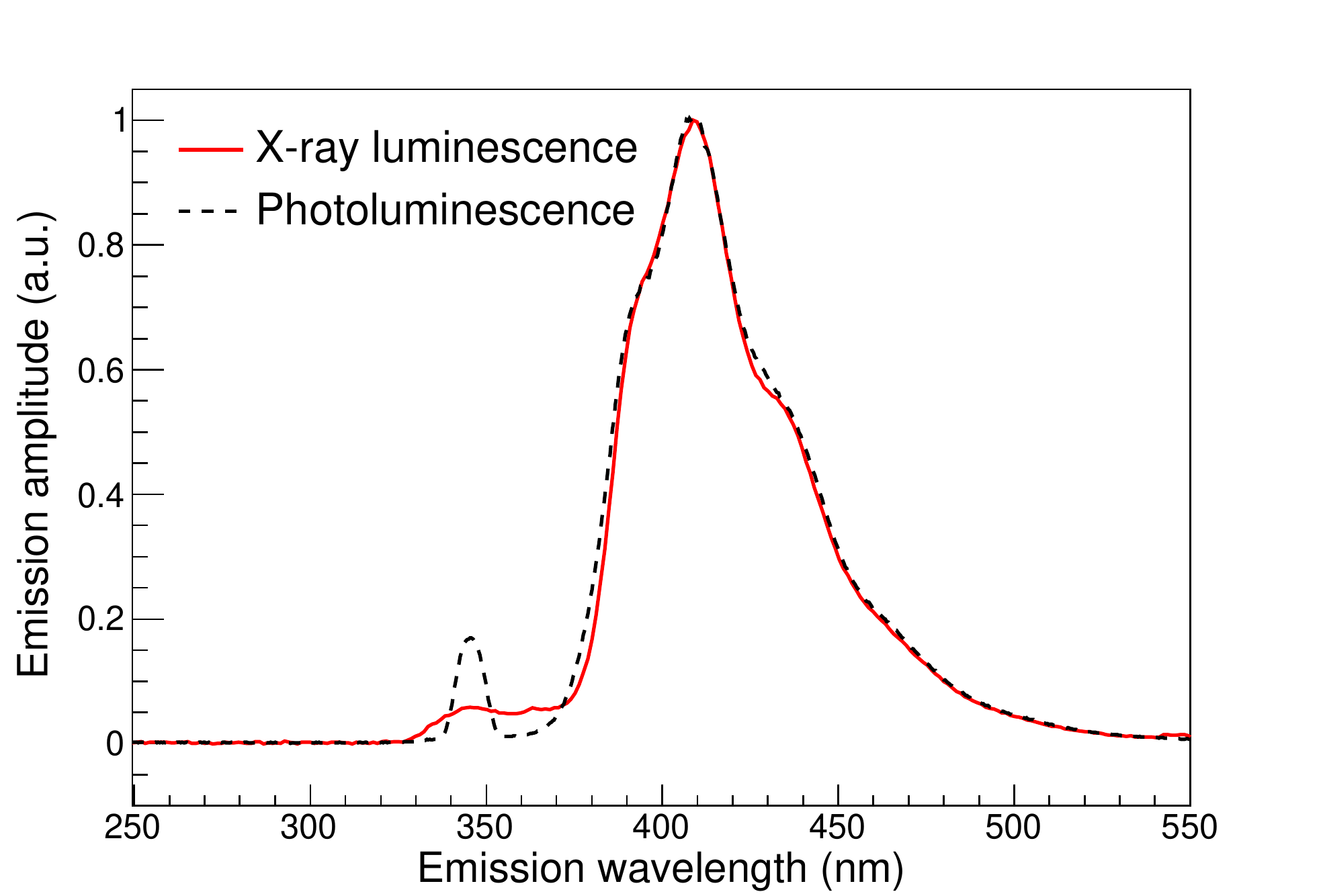}
	}
	\caption{(Color online) EJ-204 emission spectra for the (a) fresh and (b) aged samples measured via X-ray luminescence (red solid line) and photoluminescence (black dashed line). The photoluminescence spectra were obtained by averaging data from excitation wavelengths between 340-350~nm. \label{fig:emission}}
\end{figure}

Fig.~\ref{fig:emission} shows the measured photoluminescence and X-ray luminescence emission spectra of the fresh and aged EJ-204 samples. The photoluminescence spectra correspond to the average of 11~emission spectra obtained with excitation wavelengths between 340 and 350~nm in 1~nm increments. The feature centered at 345~nm corresponds to reflection of the excitation source off the sample. For the fresh sample emission spectra shown in Fig.~\ref{new-em}, the shoulder in the photoluminescence spectrum (in the region of $370-390$~nm) that is not present in the X-ray luminescence spectrum likely corresponds to excitation at the surface of the sample where the impact of self-absorption is limited~\cite{Zaitseva2015}. The aged sample, shown in Fig.~\ref{old-em}, demonstrates a relative increase in emission amplitude around $380-400$~nm when compared to the fresh sample. In addition, a shelf is observed in the X-ray luminescence spectrum of the aged sample below 380~nm. These features suggest contributions from the primary fluor in the emission spectra of the aged sample. 

\subsection{Effective quantum efficiency: $ \eta(\lambda) \epsilon_{LC}(\lambda)$}
\label{sec:EQE}

\begin{figure*}
	\subfloat[Bare APD]{
		\includegraphics[width=0.3\textwidth]{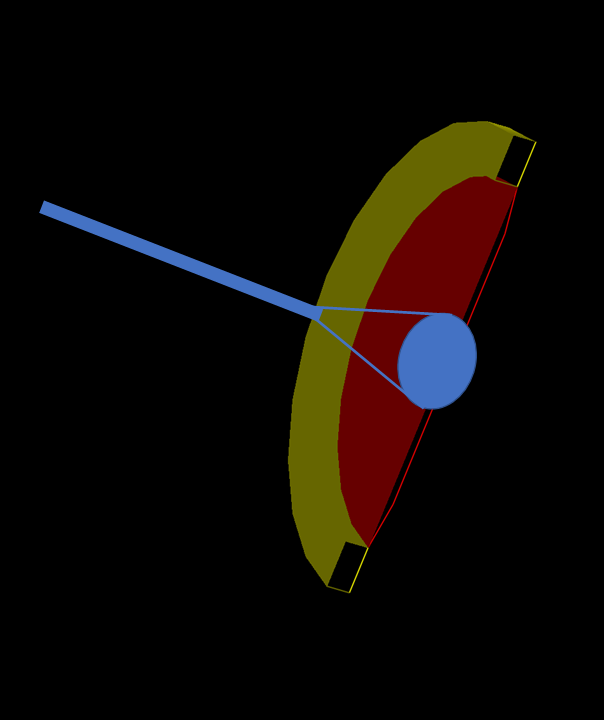}
	}\hfill
	\subfloat[Direct illumination with cap]{
		\includegraphics[width=0.3\textwidth]{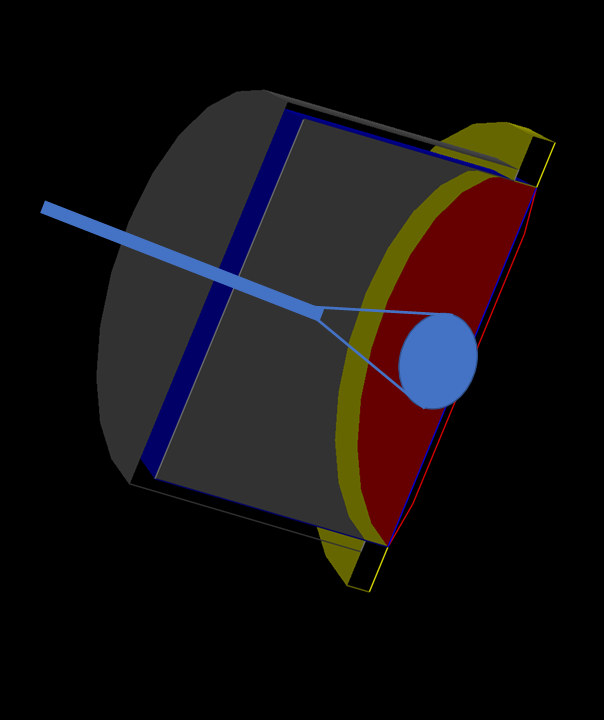}
		}\hfill
	\subfloat[Indirect illumination with cap]{
		\includegraphics[width=0.3\textwidth]{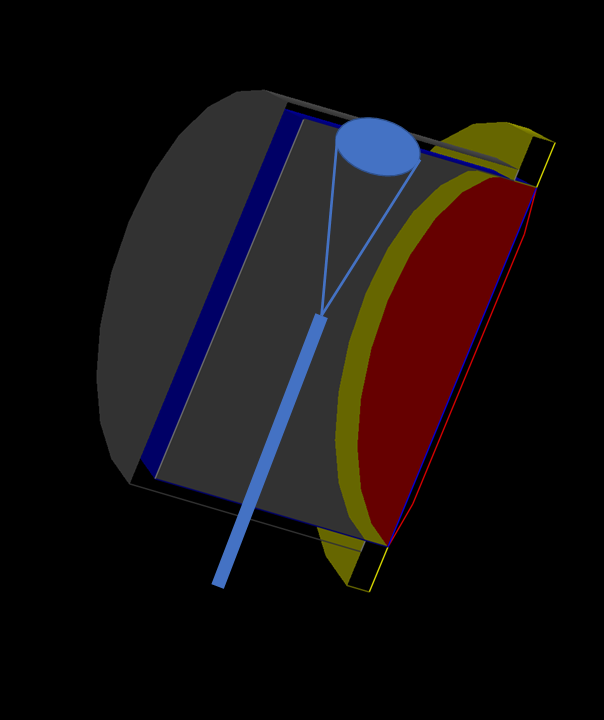}
	}
   	\caption{(Color online) Depiction of the three measurement configurations used to constrain optical model parameters to obtain the effective quantum efficiency of the APD and reflective cap. The active area of the APD is represented in red, the cannula and beam spot are shown in light blue, the copper ring of the APD is shown in yellow, and the aluminum housing is represented in gray with reflective paint shown in dark blue.}
	\label{fig:EQEconfigs}
\end{figure*}

A Geant4 optical model was benchmarked using several measurements to calculate the effective quantum efficiency of the system (i.e., the probability that a scintillation photon of wavelength $\lambda$ emitted from the scintillating volume is detected by the APD), defined as the product of the wavelength-dependent quantum efficiency of the photodetector and light collection efficiency of the housing: $ \eta(\lambda) \epsilon_{LC}(\lambda)$. A Xe lamp and excitation monochromator from a Fluorolog-3 spectrophotometer (HORIBA Jobin Yvon) served as the light source. An implantable cannula of Type CFMLC12L20 from ThorLabs Inc., consisting of an optical fiber attached to a ceramic ferrule, was used to generate a light spot sized to fit within either the active area of the APD or a S120VC photodiode power sensor from Thorlabs Inc.\ with NIST-traceable calibration. The 20-mm-long flat-cleaved cannula consisted of a $\O$200~$\mu$m core and a 0.39 numerical aperture fiber with a $\O$1.25~mm ferrule. The cannula was connected using a ThorLabs Inc.\ Optogenetics patch cable to a ThorLabs Inc.\ SM1FC2 FC/PC Fiber Adapter Plate, which was mounted in front of the spectrophotometer to collect the light from the monochromator. During illumination of the APD, the resultant current was measured using a Keithley 6517B electrometer. Additionally, the photodiode power sensor was illuminated in the same geometry to provide normalization. 
 
Measurements were performed in three configurations, illustrated in Fig.~\ref{fig:EQEconfigs}: direct illumination of the bare APD with the fiber axis perpendicular to the APD surface; direct illumination of the APD with reflective cap, with the fiber axis perpendicular to the APD surface; and indirect illumination of the APD with reflective cap, with the fiber axis parallel to the APD surface (i.e., light shone onto the side of the reflective cap). The latter two configurations were accomplished by inserting the cannula in a 1-mm-dia.\ hole bored in the reflective cap. In those cases, the ferrule was secured to a set of one-dimensional motion stages to allow for precise insertion of the cannula into the cap. 

\begin{figure}
	\center
	\includegraphics[width=0.5\textwidth]{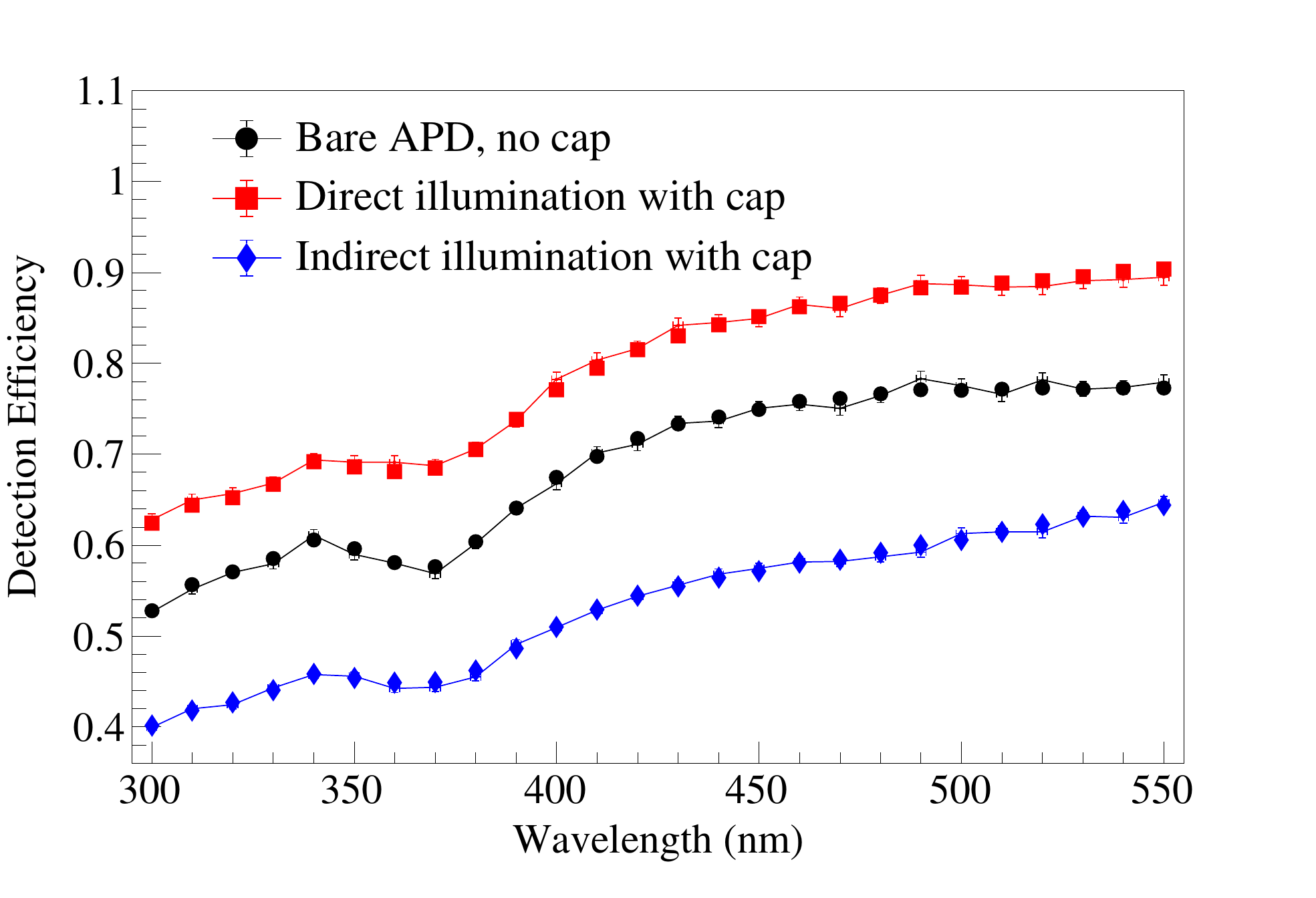}
	\caption{(Color online) Measured (data points) and simulated (error bars with lines to guide the eye) detection efficiencies for the bare APD (black circles), direct illumination of the APD with reflective cap (red squares), and indirect illumination of the APD with reflective cap (blue diamonds). The statistical uncertainty of the measured data is smaller than the data points. For the simulation output, only the statistical uncertainty is plotted. \label{fig:eqeMeasSim}}
\end{figure}

The quantum efficiency of the APD, $\eta(\lambda)$, is defined as the fraction of incident photons of wavelength, $\lambda$, that are converted into e-h pairs. This quantity was measured by impinging a well-characterized source of light onto the photodetector and determining the responsivity, $S$, defined as the ratio of the photocurrent to the incoming beam power. Then, $\eta$ is given by~\cite{Sibley2020}:
\begin{equation}
\eta = \frac{hcS}{e\lambda},
\end{equation}
where $h$ is Planck's constant, $c$ is the speed of light in vacuum, and $e$ is the charge of the electron. The manufacturer-provided responsivity of the NIST-calibrated power sensor was used to determine the incoming beam power. Then, the responsivity of the APD was obtained using the measured APD current for a given input beam power as a function of wavelength. The detection efficiencies for direct and indirect illumination of the APD with reflective cap were similarly determined. The measured data points are provided in Fig.~\ref{fig:eqeMeasSim}. 

Geant4~\cite{Geant4} optical simulations were then performed using the UNIFIED model~\cite{Levin1996,Nayar1991} to obtain the bare APD quantum efficiency and the detection efficiencies for the direct and indirect illumination configurations with the reflective cap. The painted surface of the cap and active area of the APD were treated as optical surfaces upon which optical photons could either be reflected or absorbed. Following the manufacturer specifications, the paint was modeled as a pure Lambertian reflector~\cite{labsphere}. In the case of the APD, pure specular spike reflectance was assumed~\cite{Wagenpfeil_2021}. For each wavelength between 300 and 550~nm in 10~nm increments, a $\chi^2$ minimization was performed between the measured data for each of the three configurations and the simulation output, where the reflectivity of the paint and APD surface as well as the efficiency of the APD (i.e., the probability for a photon absorbed at the surface of the APD to lead to a detectable signal) were treated as free parameters.  

\begin{figure}
	\center
	\includegraphics[width=0.5\textwidth]{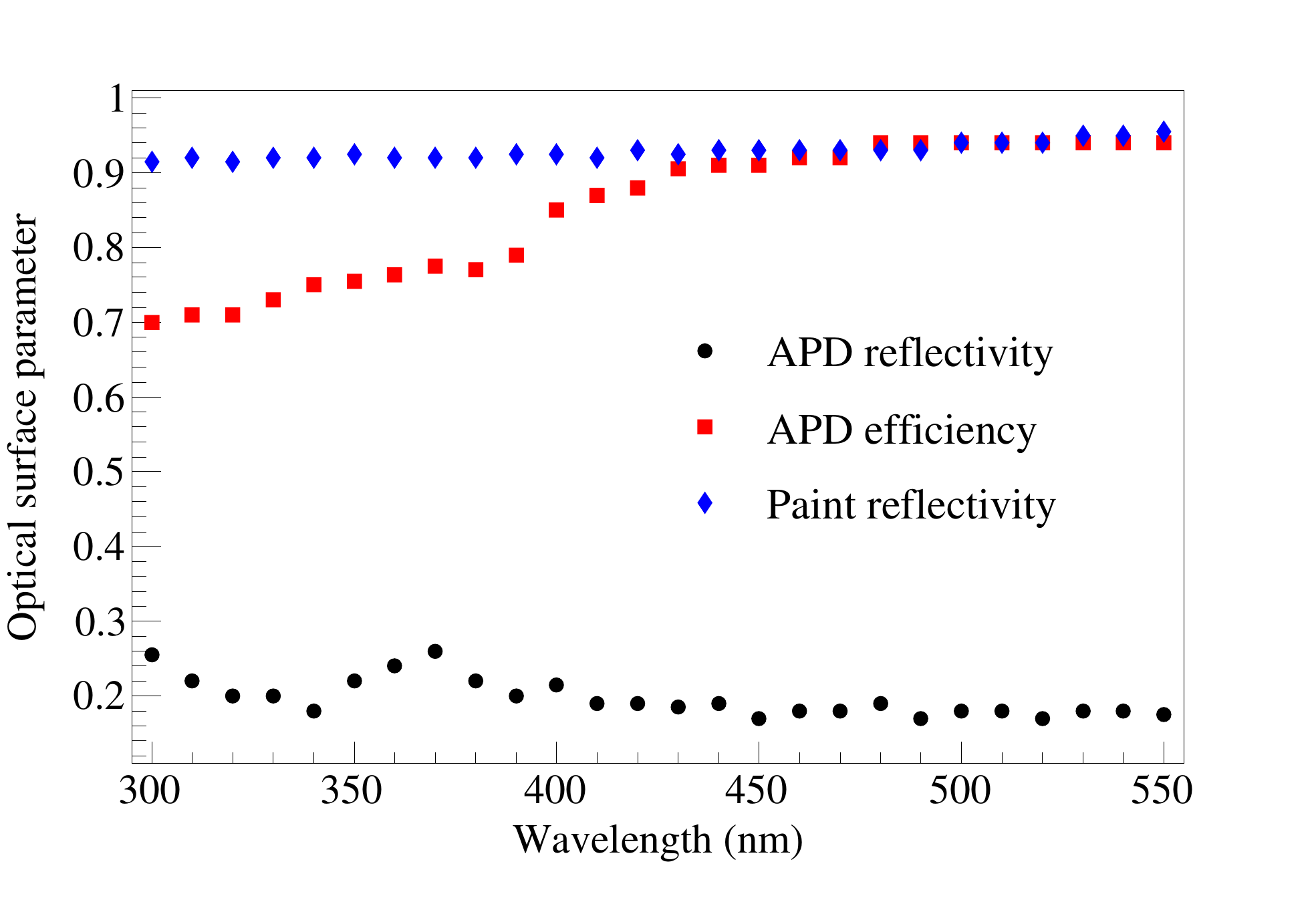}
	\caption{(Color online) Best-fit optical parameters determined using a $\chi^2$ minimization between the measured detection efficiency of the system in three configurations and the Geant4 simulation output.\label{fig:bestOpticalParams}}
\end{figure}

Fig.~\ref{fig:eqeMeasSim} shows the measured data along with the Geant4 simulation output for the bare APD (black circles), direct illumination of the APD with reflective cap (red squares), and indirect illumination of the APD with reflective cap (blue diamonds). The agreement between the simulation and measured data is within the statistical uncertainty of the simulation ($1\%$) for all of the measured data points. The best-fit optical parameters are shown in Fig.~\ref{fig:bestOpticalParams}. 

To propagate covariance arising from the optical parameter uncertainties, the effective quantum efficiency and associated uncertainty were obtained via Monte Carlo sampling of the optical parameter curves in Fig.~\ref{fig:bestOpticalParams} with 10,000 trials. Specifically, each of the optical parameter curves was sampled from a normal distribution with a standard deviation given by its estimated uncertainty. For each trial, $i$, four Geant4 simulations were performed: direct illumination of the bare APD, direct illumination of the APD with reflective cap, indirect illumination of the APD with reflective cap (i.e., the three configurations illustrated in Fig.~\ref{fig:EQEconfigs}), and the APD with reflective cap and scintillator treated as a uniform photon source over the full scintillating volume. The interface between the scintillator and the air was modeled as a ground Lambertian surface and allowed for reflection or refraction following Snell's Law. The $\chi^2$ per degrees of freedom, $\widetilde{\chi}^2$, was computed using the first three simulations to characterize the goodness-of-fit between the measured detection efficiencies provided in Fig.~\ref{fig:eqeMeasSim} and the simulation output. The effective quantum efficiency was computed from the fourth, determined as the wavelength-dependent detection efficiency for the APD with reflective cap in response to a uniform, isotropic photon source over the volume of the scintillator. The mean and standard deviation of the effective quantum efficiency were then calculated over the ensemble of trials, weighted by the Birge factor, $w_i$, given by~\cite{Birge1932}:
 \begin{equation}
w_i = \frac{1}{\sqrt{{\widetilde{\chi}_i^2}}}.
\end{equation}
The spectrum-averaged effective quantum efficiency and uncertainty associated with the optical parameters was determined to be $0.570 \pm 0.011$ and $0.559 \pm 0.011$ for the fresh and aged samples, respectively. The spectrum integration was performed using the measured X-ray luminescence emission spectra for each sample, as it takes into account absorption and reemission in the volume of the scintillator and is thus more representative of the emission spectrum of the medium when excited by $\gamma$ rays.

Further sources of systematic uncertainty were identified. The parameter characterizing the surface roughness of the scintillator in the Geant4 simulations was varied over its full range to investigate the impact on the effective quantum efficiency and a maximum deviation of $0.3\%$ was observed. The stability of the fluorometer lamp along with the power sensor were estimated by comparing three separate measurements, leading to a $0.5\%$ uncertainty. The impact of the distance between the cannula tip and the APD was investigated by repeating the measurement with variable insertion depths and a $1.2\%$ uncertainty was observed. For each of these measurements, the calculated uncertainty was weighted by the normalized emission spectrum of the scintillator. The quoted manufacturer uncertainty for the NIST-traceable calibration of the power sensor was $5\%$~\cite{thorlabsPowerSensor}. The spectrum-averaged effective quantum efficiency was calculated for repeated measurements of the X-ray luminescence spectrum, and the values differed by 0.6\%. Finally, a 1-mm uncertainty in the position of the scintillator relative to the back of the cap corresponded to a $0.7\%$ increase in light collected, as estimated using a Geant4 simulation. 

Due to the difference in refractive index between the scintillator and air, total internal reflection within the scintillator is possible, which can result in light trapping~\cite{BirksBook-Ch5}. Given the small size of the scintillating volume employed in this work and the relatively long attenuation length of the material \cite{Eljen-EJ-20x-specs}, the amount of scintillation photons trapped within the scintillator is expected to be small. This effect was studied in simulation space using the best-fit optical parameters and the number of photons absorbed in the EJ-204 scintillator corresponded to 0.25$\%$ of the total photons.

\subsection{Number of electron-hole pairs created: $N_{e{\text -}h}$}
\label{CE}

\begin{figure}
	\center
	\includegraphics[width=0.5\textwidth]{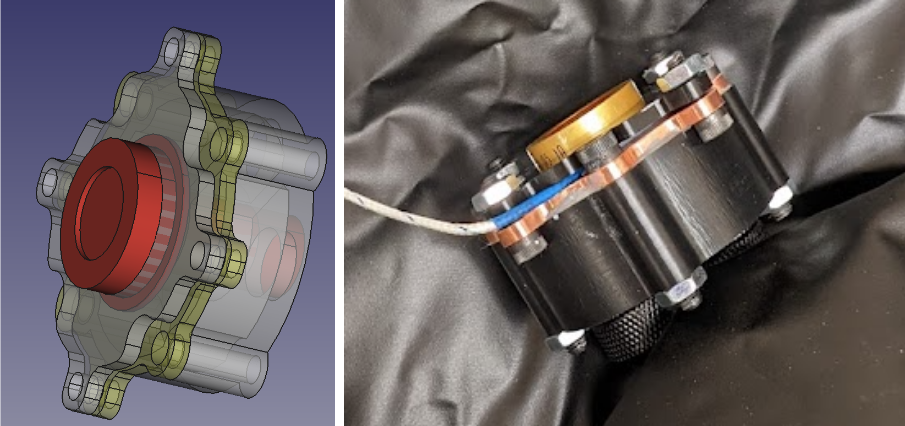}
	\caption{(Color online) Three-dimensional model (left) and photograph (right) of the APD coupled to a cooling block.\label{fig:coolingBlock}}
\end{figure}

The APD was employed to measure the e-h pair yield of the scintillation response. The detector featured a 16-mm-diameter active area with enhanced quantum efficiency in the ultraviolet region. Temperature control was managed using a cooling block designed to work with a closed-loop chiller. The APD was sandwiched onto a copper plate exposed to a recirculating water loop, illustrated in Fig.~\ref{fig:coolingBlock}. Thermal coupling between the APD and the copper block was achieved using a conductive silicone pad. The cooling water loop was maintained to $12 \pm 0.1^\circ$C using a Julabo F25 MD recirculating chiller.

The APD signal was read out using an Ortec Model 142PC preamplifier fed into a modified Texas Instruments THS4303EVM amplifier with a gain of 40, powered by a Hewlett-Packard 6235A Triple Output Power Supply. The bias voltage was maintained at 1675~V. The signal was fed to a CAEN V1730 digitizer and processed using the DPP-PHA firmware to extract pulse amplitude values. For digital signal processing, the DPP-PHA trapezoidal filter was applied, which used a 696~ns rise time with a 600~ns flat top and a 64\% flat top delay~\cite{DPP-PHA}. The peak average was evaluated using 4 samples.

The APD, preamplifier, and amplifier were placed in a light-tight housing constructed from an electronically shielded enclosure to reduce outside sources of noise. A Peltier cooling assembly was added to provide dehumidification and ambient temperature control for the enclosure. The response of the 14.4~keV $\gamma$ ray from the $^{57}$Co source was used to demonstrate gain stability to within 1.2$\%$ over the duration of the data collection period.

\begin{figure}
	\center
	\includegraphics[width=0.47\textwidth]{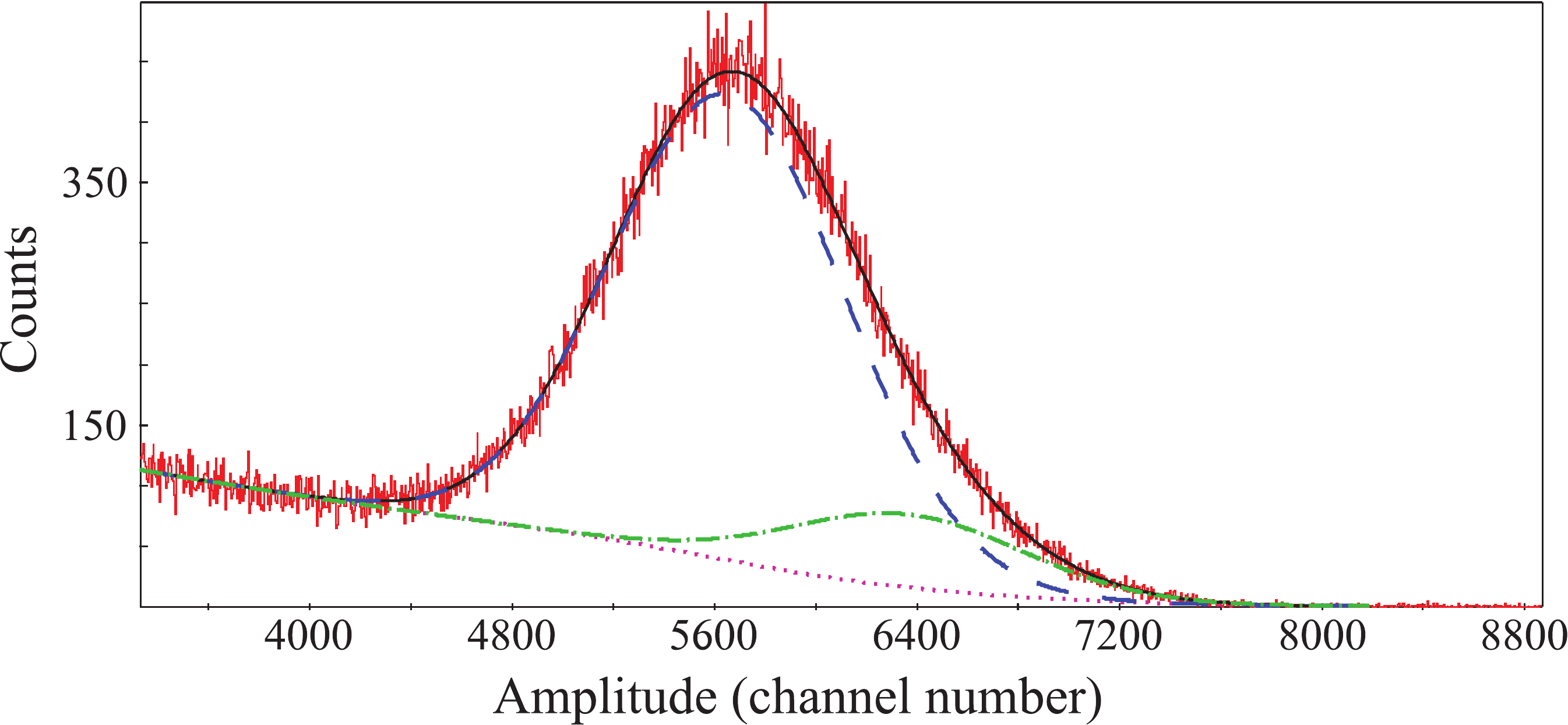}
	\caption{(Color online) Energy spectrum of X rays from the $^{109}$Cd source (red) along with a fit of the data (black line). The fit was composed of two Gaussian distributions (dashed blue and dot-dashed green) and a second-order polynomial background term (dotted pink). \label{fig:Cd}}
\end{figure}

Measurements were performed with $^{57}$Co and $^{109}$Cd sources positioned several cm from the front face of the bare APD. The isolated 14.4~keV $\gamma$ ray from the $^{57}$Co source (measured at the start and end of the data collection period) was fit with a Gaussian distribution to obtain an average centroid of $3635.0 \pm 30.4$ analog-to-digital-converter (ADC) channels. The measured $^{109}$Cd X-ray spectrum is shown in Fig.~\ref{fig:Cd}. In this energy region, the source produces five unresolved X rays (see Table~\ref{tab:cdxrays}). Spectral decomposition was accomplished by fitting two superposed Gaussian distributions, where the transitions at 21.99~keV and 22.163~keV were characterized as a doublet and the transitions at 24.912~keV, 24.943~keV, and 25.455~keV were treated as a broad multiplet. A second-order polynomial background term was included to characterize the energy-dependent noise contribution, and the relative area of the fitted Gaussian distributions was consistent with known X-ray emission probabilities. The centroid of the doublet was determined to be $5650.4 \pm 8.4$ ADC channels, which was assigned a weighted-average energy of 22.103 keV. 

\begin{table}
	\caption{Energy and intensity of X rays arising from the $\beta$-decay of Cd-109~\cite{Kumar2016}.}
	\label{tab:cdxrays}
	\centering
	\begin{tabular}{cc}
		\hline
		Energy (keV)   & Intensity (\%) \\
		\hline
		21.99 & 29.8 \\ 
		22.163 & 56.1 \\ 
		24.912 & 4.80 \\ 
		24.943 & 9.3 \\ 
		25.455 & 2.31 \\
		\hline
	\end{tabular}
\end{table}

\begin{figure}
	\center
	\includegraphics[width=0.47\textwidth]{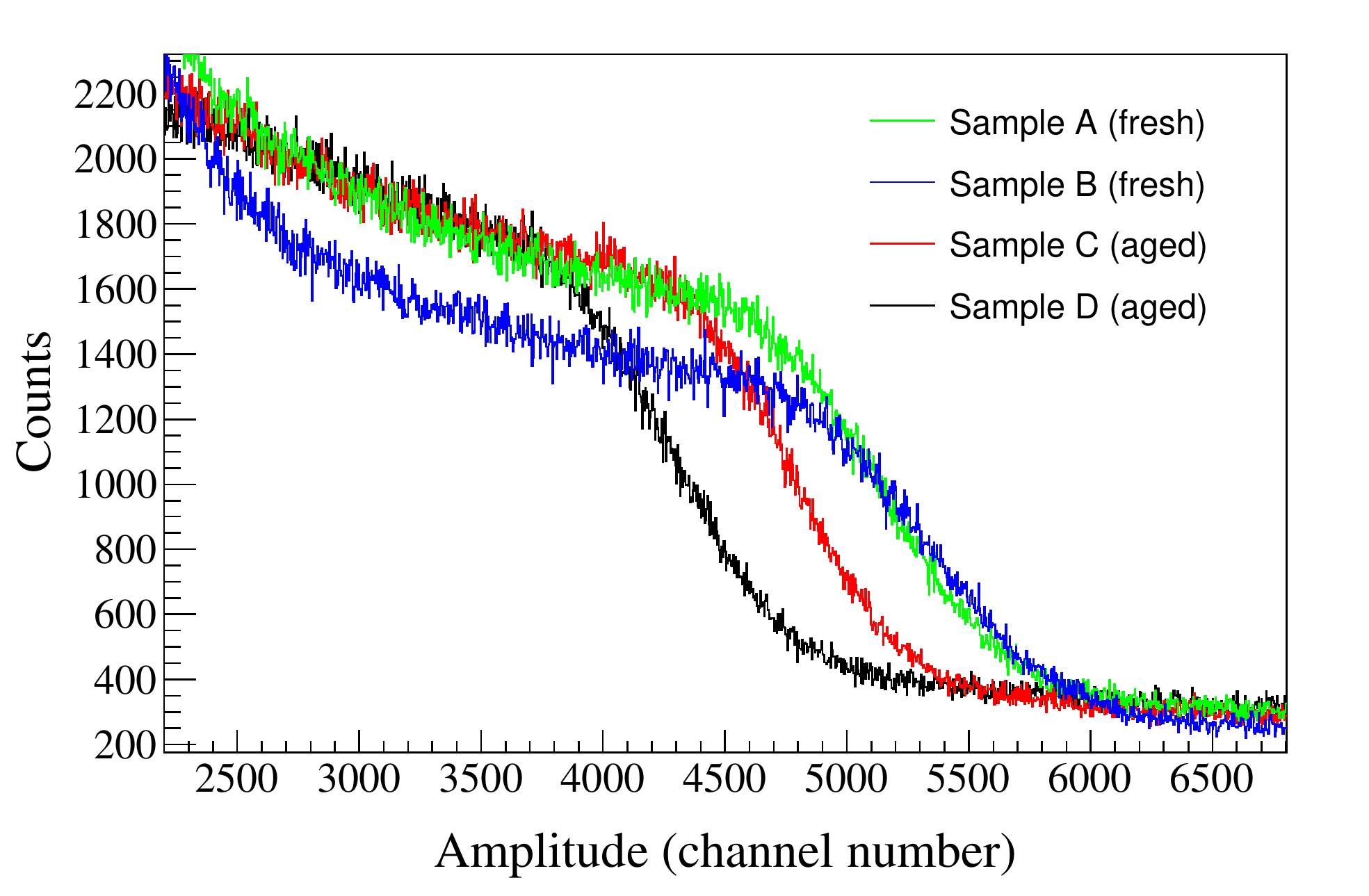}
	\caption{(Color online) Compton edge spectra for a $^{22}$Na source incident on the four EJ-204 samples examined.\label{fig:compton}}
\end{figure}

Four EJ-204 samples were examined, each 15-mm-dia.\ x 5-mm-h.\ right circular cylinders. Two samples (A and B) were manufactured approximately four months prior and stored in high-purity nitrogen. Two samples (C and D) were aged approximately eleven years and stored in ambient air. Each scintillator was positioned within the reflective cap which was mounted on the APD with an air gap between the scintillator and the APD surface. Pulse amplitude spectra were obtained using a $^{22}$Na source placed atop the APD housing to determine the position of the Compton edge of the 1274.5~keV $\gamma$ ray. The response of the four EJ-204 samples to the $^{22}$Na source is shown in Fig.~\ref{fig:compton}. 

\begin{figure}
	\center
	\includegraphics[width=0.47\textwidth]{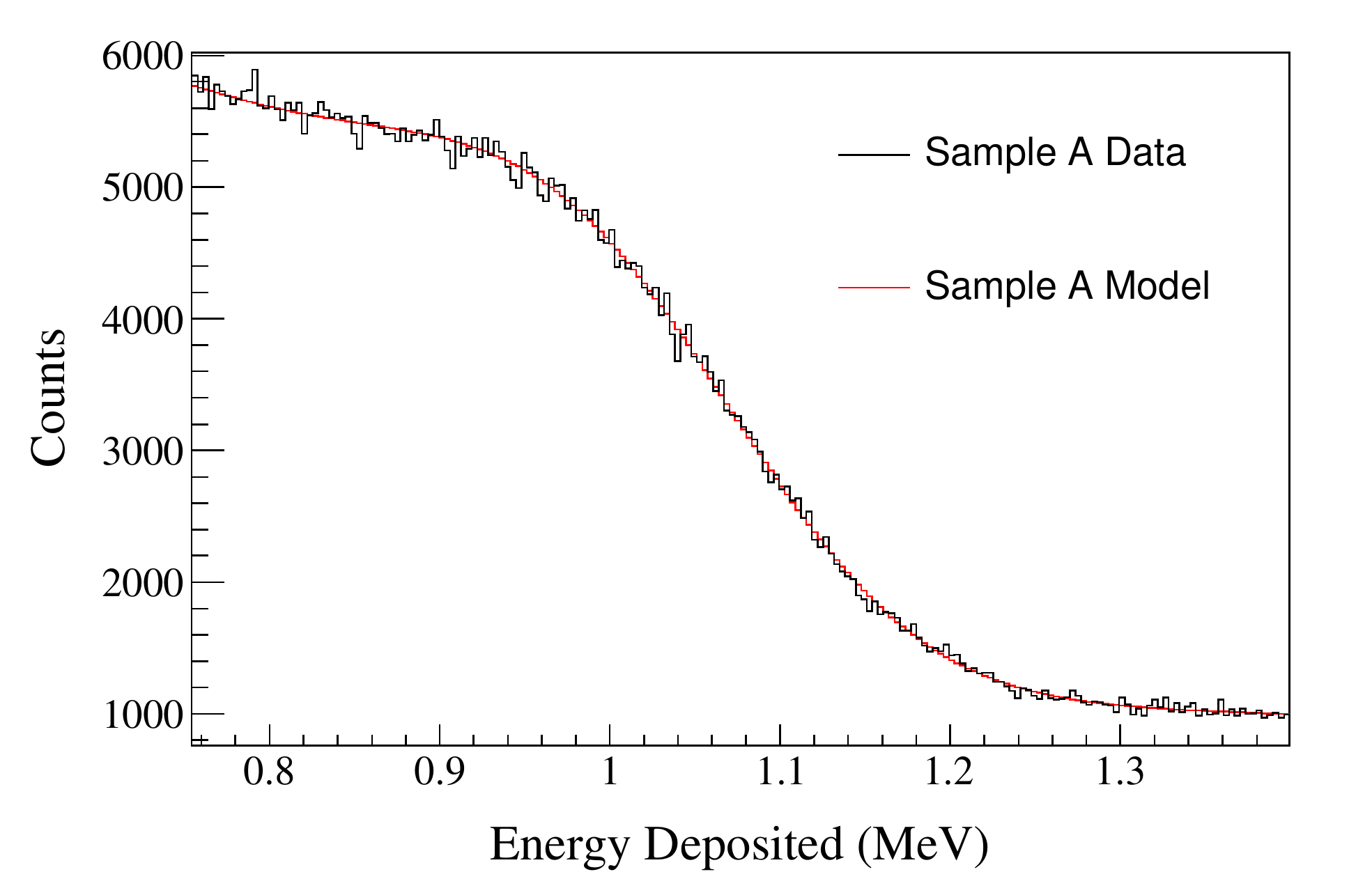}
	\caption{(Color online) Compton edge spectrum for EJ-204 Sample A (black) fit with a simulated energy-deposition spectrum convolved with an energy-dependent resolution function (red). \label{fig:comptonfit}}
\end{figure}

The position of the Compton edge was determined by fitting the experimental data with a simulated energy deposition spectrum convolved with an energy-dependent resolution function plus a power law background term. The Geant4 toolkit was used to simulate the scintillator response as described in~\cite{Brown2018}. In this case, the resolution function of the APD described by Moszy\'{n}ski, et al.\ was implemented~\cite{Moszynski2000}:
\begin{equation}
(\Delta E/E)^2 = (\delta_{\text{sc}})^2 + (\Delta N/N_{e{\text -}h})^2 + (\Delta_{\text{noise}}/N_{e{\text -}h})^2,
\end{equation}
where $\Delta E/E$ is the energy resolution, $\delta_{\text{sc}}$ is the intrinsic resolution of the scintillator, $\Delta N/N_{e{\text -}h}$ is the statistical contribution, and $\Delta_{\text{noise}}/N_{e{\text -}h}$ is the APD noise contribution. A representative fit of the Compton edge of the $^{22}$Na $\gamma$ ray for Sample A is shown in Fig.~\ref{fig:comptonfit}. 

Using the mean e-h pair formation energy in Si of 3.67 eV~\cite{Pehl1968}, the $^{57}$Co and $^{109}$Cd calibrations were quadratically interpolated to determine the number of electron-hole pairs produced at each of the $^{22}$Na Compton edges. Table~\ref{tab:numpairs} provides the number of e-h pairs determined for a 1061.7~keV electron energy deposition corresponding to the Compton edge of the 1274.5~keV $\gamma$ ray for the four scintillator samples. For the system gain and operating temperature, gain nonlinearity between the $\gamma$/X rays and visible light pulses was estimated to be 0.2\%~\cite{Fernandes2007}.

\begin{table}
	\caption{Channel number and number of e-h pairs corresponding to the Compton edge of the 1274.5~keV $\gamma$ ray for the EJ-204 samples.}
	\label{tab:numpairs}
	\centering
	\begin{tabular}{ccc}
		\hline
		Sample   & Channel Number & Number of e-h pairs  \\
		\hline
		A & 5269 & 5630 $\pm$ 178 \\
		B & 5091 & 5447 $\pm$ 169 \\
		C & 4228 & 4549 $\pm$ 130 \\
		D & 4728 & 5071 $\pm$ 152 \\
		\hline
	\end{tabular}
\end{table}

A Monte Carlo simulation was performed using Geant4~\cite{Geant4} to investigate the potential for bias associated with partial energy deposition owing to the small size of the scintillating volume~\cite{Weldon2020}. The geometry consisted of the APD and the cylindrical scintillator contained within the aluminum housing. A $^{22}$Na source was simulated via 1274.5~keV $\gamma$ rays emitted isotropically orthogonal to the midpoint of the side of the scintillator and at a distance of several cm. The source was moved off axis relative to the midpoint at 1~cm, 1.5~cm, and 2~cm and the simulation was repeated. The energy deposition spectrum and associated fraction of events resulting in recoil electron escape were determined for each source position. Although 22-24\% of events in the energy deposition spectra resulted in electrons escaping the scintillator prior to full energy deposition, the impact of recoil electron escape on the determination of the position of the Compton edge was negligible. 

\section{Results and Discussion}
\label{sec:Results}

The number of e-h pairs, given in Table~\ref{tab:numpairs}, was divided by the spectrum-averaged effective quantum efficiency to provide the photon yield for a 1061.7~keV electron energy deposition. Assuming local electron light linearity,\footnote{Given the known nonproportionality of the electron response of the EJ-200 plastic scintillator from Eljen Technology~\cite{Payne2011} and its commercial equivalent BC-408~\cite{Nassalski2008, Swiderski2012}, which share the same PVT polymer base as EJ-204, the absolute light yields reported in this work are not expected to be linear across a broad energy range. Extrapolation of the reported values should take into account the electron light yield nonproportionality.} the absolute light yields in photons per MeV of the four EJ-204 samples examined in this work are provided in Table~\ref{tab:results}. The sources of uncertainty for the absolute light yield measurement are detailed in Table \ref{tab:unc}. The light yields determined for the fresh samples, A and B, agree within one standard deviation, with an average value of $9100 \pm 400$ photons per MeV. 

The fresh samples exhibited a mean absolute light yield that is lower than the manufacturer-provided value of 10,400 photons per MeV \cite{Eljen-EJ-20x-specs}. The relatively lower EJ-204 absolute light yield observed in this work is not unexpected, as the value provided in the materials specification sheet was determined relative to an EJ-200 sample with an absolute light yield set at 10,000 photons per MeV~\cite{Hurlbut2018}. The EJ-200 photon standard was guided by previous absolute light yield measurements of EJ-208 and EJ-212 by Holl et al~\cite{Holl1988}, which are expected to be biased high given the use of the half height of the $\gamma$-ray distribution to characterize the position of the Compton edge~\cite{Dietze1982}. 

The aged samples, C and D, exhibited significantly lower absolute light yields compared to the manufacturer-provided value by approximately 26\% and 18\%, respectively, despite being optically clear. This may be due to a number of factors. Plastic scintillators formulated using a PVT base have been shown to degrade as a result of exposure to oxygen and ultraviolet light~\cite{Weir1972}. Significant recent work to understand the root cause of plastic scintillator degradation also points to the effects of temperature and humidity cycling~\cite{Janos2020,Myllenbeck2020,Lance2020,Kouzes2020}, which have been shown to impact samples to differing degrees and, in some cases, lead to a reduction in light yield in the absence of fogging~\cite{Loyd2020}. 

\begin{table}
	\caption{Absolute light yield of EJ-204. Samples A and B were aged approximately four months and stored in high-purity nitrogen, while Samples C and D were aged approximately eleven years and exposed to ambient air.}
	\label{tab:results}
	\centering
	\begin{tabular}{cc}
		\hline
		Sample   & Photons per MeV  \\
		\hline
		A & 9300 $\pm$ 600  \\
		B & 9000 $\pm$ 600 \\
		C & 7700 $\pm$ 500 \\
		D & 8500 $\pm$ 500 \\
		\hline
	\end{tabular}
\end{table}

\begin{table*}[!htp]

	\caption{Sources of uncertainty for each component of the absolute light yield measurement as well as the combined total uncertainty.}
	\label{tab:unc}
	\centering
	\begin{threeparttable}
	\footnotesize
	\begin{tabular}{ccc}
		\hline
		Quantity   &  Description & Uncertainty (\%)  \\
		\hline
		Compton edge (e-h pairs) & $\gamma$/X-ray calibration & 3.1\tnote{a} \\
		 & Compton edge determination & 0.02 \\
		 & Non-linearity between $\gamma$/X-rays and visible light gain & 0.2 \\
		 & \textit{Total} & \textit{3.1} \\
		\hline
		Spectrum-weighted effective quantum efficiency& Model error & 2.0 \\
		 & Lamp stability & 0.5 \\
		 & Cannula positioning & 1.2 \\ 
		 &Power sensor calibration&5\\
		 & Emission spectrum & 0.6 \\
		 & Scintillator position & 0.7 \\
		 & Light trapping & 0.25 \\
		 & \textit{Total} & \textit{5.6} \\
		\hline
		\textbf{Absolute light yield} & \textbf{Total} & \textbf{6.4\%}\\
		\hline
	\end{tabular}
	\begin{tablenotes}
	\item[a] The uncertainty on the calibration performed using low energy $\gamma$-ray and X-ray sources ranged from 2.9 to 3.1\% for the various samples.
	\end{tablenotes}
	\end{threeparttable}
\end{table*}

\section{Summary}
\label{sec:Summary}

The absolute light yield of four EJ-204 plastic organic scintillator samples was examined, two fresh samples fabricated four months prior and stored in high-purity nitrogen and two aged samples fabricated approximately eleven years prior and exposed to ambient air. The average absolute light yield of the fresh samples was determined to be $9100 \pm 400$ photons per MeV, approximately 13\% lower than the manufacturer-provided value of 10,400 photons per MeV. The aged samples exhibited significantly decreased absolute light yield (up to 26\% lower than the manufacturer specification). While potential mechanisms for aging of the EJ-204 plastic scintillators are known, the relative importance of the different factors are not fully understood, and experimenters should use caution when employing this medium in photon counting applications or as a reference standard for future absolute light yield measurements. This work provides the first direct measurement of the absolute light yield of EJ-204 and contributes to the body of literature on environmental aging of plastic scintillators.  

\section*{Acknowledgments}
The authors are grateful for discussions and expert advice from Chuck Hurlbut and Erik Brubaker. Weronika Wolszczak is recognized for her support with the emission spectrum measurements. This work was performed under the auspices of the U.S. Department of Energy by Lawrence Berkeley National Laboratory under Contract DE-AC02-05CH11231 and the U.S. Department of Energy National Nuclear Security Administration through the Nuclear Science and Security Consortium under Award Nos.\ DE-NA0003180 and DE-NA0003996. The project was funded by the U.S. Department of Energy, National Nuclear Security Administration, Office of Defense Nuclear Nonproliferation Research and Development (DNN R\&D).

\bibliographystyle{elsarticle-num}
\bibliography{./AbsLY.bib}

\end{document}